\documentclass[
    ,final            
    ,numberedheadings 
    ,cmfonts          
  ]
  {aipproc}

\layoutstyle{6x9}

\begin{document}

\title{Partial-Wave Analysis of Single-Pion Production Reactions}


\author{R.A.~Arndt}
{address={Center for Nuclear Studies, Department of Physics,
           The George Washington University, Washington, D.C. 
           20052, USA}}
\author{W.J.~Briscoe}
{address={Center for Nuclear Studies, Department of Physics,
           The George Washington University, Washington, D.C.
           20052, USA}}
\author{I.I.~Strakovsky}
{address={Center for Nuclear Studies, Department of Physics,
           The George Washington University, Washington, D.C.
           20052, USA}} 
\author{R.L.~Workman}
{address={Center for Nuclear Studies, Department of Physics,
           The George Washington University, Washington, D.C. 
           20052, USA}}

\begin{abstract}
We present an overview of our efforts to analyze pion-nucleon
elastic scattering data, along with data from related photo- 
and electroproduction reactions, in order to study the baryon
spectrum. We then focus on the $\Delta(1232)$ resonance.  
Fits to pion photo- and electroproduction data have been used 
to extract values for the $R_{EM}$ = E2/M1 and $R_{SM}$ = S2/M1 
ratios as functions of $Q^2$.  These results are compared 
to other recent determinations.
\end{abstract}

\maketitle


Many of our fits to scattering data have been motivated by 
ongoing studies of the $N^\ast$ properties~\cite{PDG}. Most 
of these require, as input, amplitudes extracted from 
\begin{figure}[htbp]
      \includegraphics[height=0.4\textwidth, angle=90]{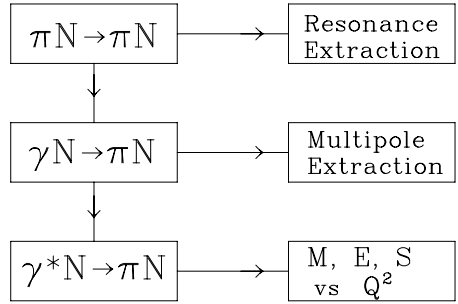}
      \caption{$\Delta$ multipoles from analysis
      of scattering data $-$ a Road Map.} \label{fig:g1}
\end{figure}
elastic pion-nucleon scattering data~\cite{gw_pin}. Our 
pion photoproduction multipoles are determined using a 
K-matrix formalism, based upon pion-nucleon partial-wave 
amplitudes~\cite{gw_photo}.  The electroproduction 
analysis is similarly anchored to our $Q^2 = 0$ 
photoproduction results, with additional factors intended 
to account for the $Q^2$ variation. This relationship is 
diagrammed in Fig.~\ref{fig:g1}.  Most of what we discuss 
\begin{figure}[htbp]
      \includegraphics[height=0.4\textwidth, angle=90]{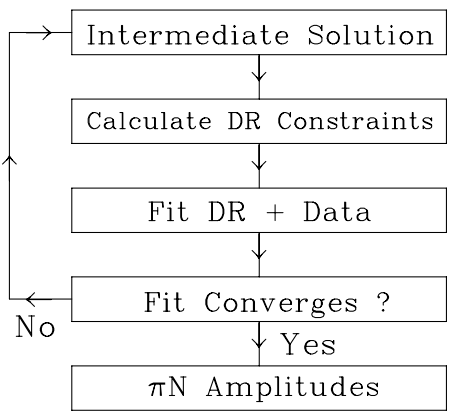}
      \caption{$\pi N$ analysis flow chart.} \label{fig:g2}
\end{figure}
here is confined to an energy region covering the 
$\Delta(1232)$ resonance.  Problems associated with the 
opening of channels beyond $\pi N$~\cite{formal} are 
avoided.

\begin{figure}[htbp]
      \includegraphics[height=0.5\textwidth, angle=90]{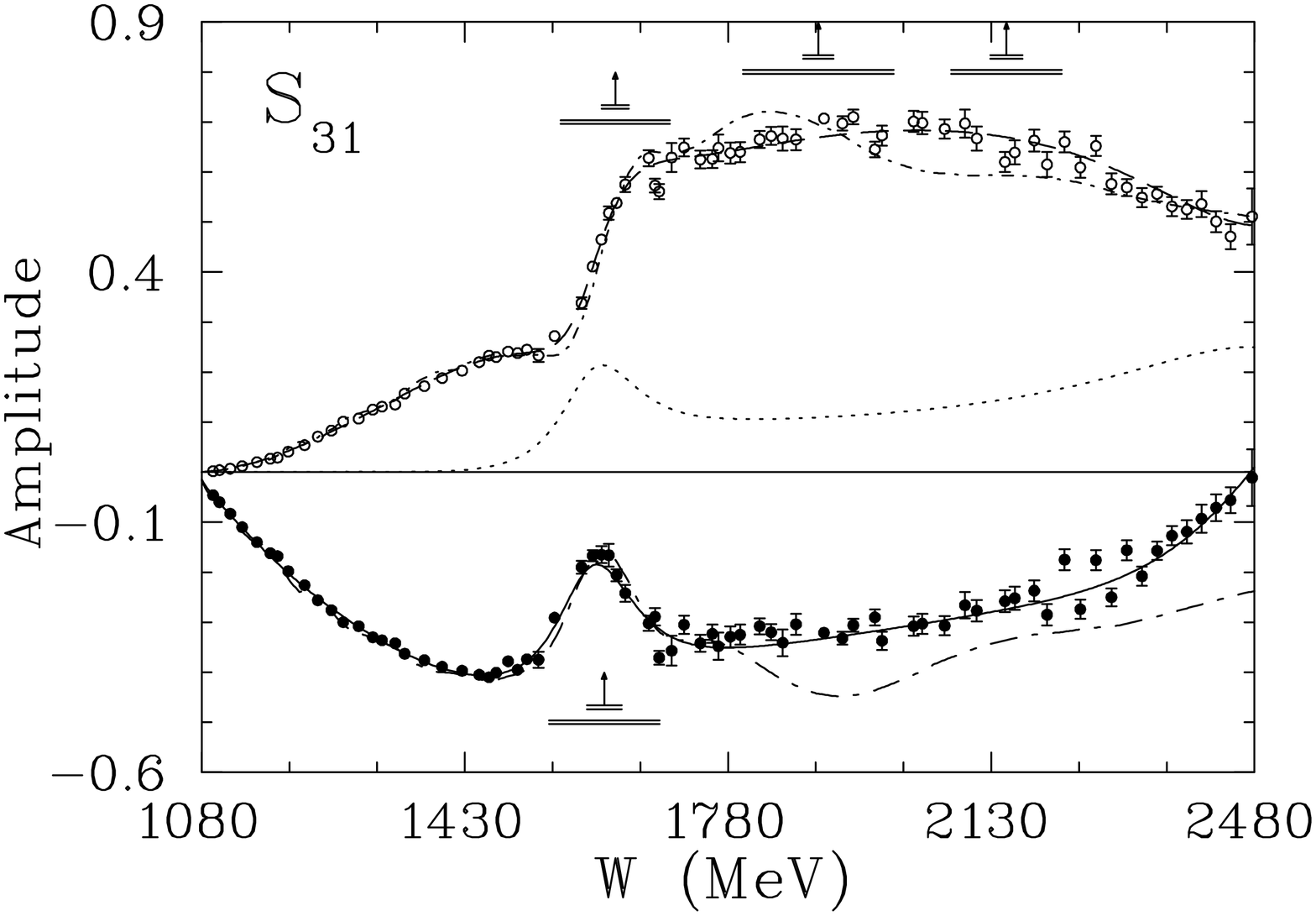}\hfill
      \includegraphics[height=0.5\textwidth, angle=90]{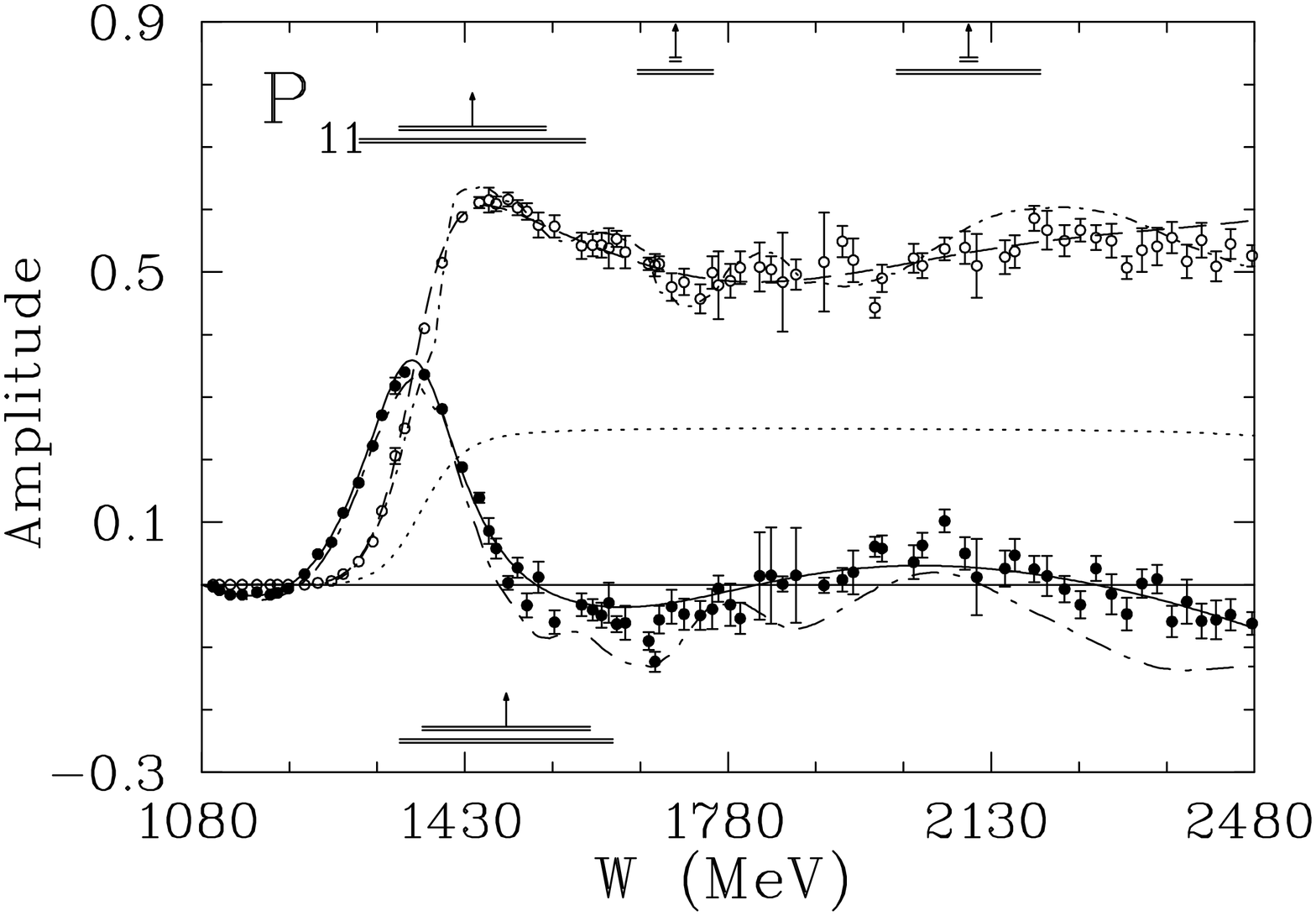}
      \caption{Partial-wave amplitudes $S_{31}$ and $P_{11}$ 
      from $T_{\pi}$ = 0 to 2.6~GeV.  Solid (dashed) curves 
      give the real (imaginary) parts of amplitudes 
      corresponding to the SP06 solution~\protect\cite{gw_pin}.  
      The real (imaginary) parts of single-energy solutions 
      are plotted as filled (open) circles.  The dotted curve 
      gives the unitarity limit ($ImT - T^{\ast}T$) from SP06.  
      The Karlsruhe KA84 solution~\protect\cite{KH} is 
      plotted with long dash-dotted (real part) and short 
      dash-dotted (imaginary part) lines.  All amplitudes 
      are dimensionless.  Vertical arrows indicate resonance 
      $W_R$ values and horizontal bars show full $\Gamma$ and
      partial $\Gamma_{\pi N}$ widths.  The lower BW 
      resonance symbols are associated with the SP06 values; 
      upper symbols give RPP~\protect\cite{PDG} values.}
      \label{fig:g3}
\end{figure}

\begin{figure}[htbp]
      \includegraphics[height=0.6\textwidth, angle=90]{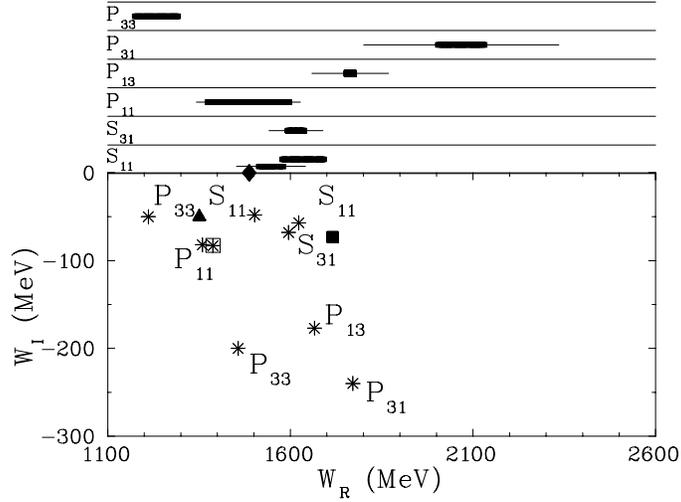}
      \caption{Comparison of complex plane (bottom panel) 
      and Breit-Wigner (top panel) parameters for 
      resonances found in the SP06 
      solution~\protect\cite{gw_pin}.  Plotted are the 
      result for S- and P-wave resonances.  Complex 
      plane poles are shown as stars (the boxed star 
      denotes a second-sheet pole).  $W_R$ and $W_I$ 
      give real and imaginary parts of the 
      center-of-mass energy.  The full ($\pi N$ partial) 
      widths are denoted by thin (thick) bars for each 
      resonance.  The branch point for $\pi\Delta(1232)$, 
      1350 - i50~MeV, is represented as a solid triangle.  
      The branch points for $\eta N$, 1487 -i0~MeV, and 
      $\rho N$, 1715 - i73~MeV, thresholds are shown as 
      a solid diamond and solid square, respectively.}
      \label{fig:g4}
\end{figure}

The pion-nucleon amplitudes are determined through a fit 
to elastic scattering, charge-exchange, and $\eta N$ 
production data, constrained to satisfy forward and fixed-t 
dispersion relations. A flowchart for this procedure is 
given in Fig.~\ref{fig:g2}. Two representative results for 
the partial-wave amplitudes are given in Fig.~\ref{fig:g3}, 
which shows that the extraction of resonance contributions 
is difficult for most states. A search of the complex 
energy plane finds poles which are often far from the 
physical axis, as shown in Fig.~\ref{fig:g4}, making a 
simple Breit-Wigner
parametrization questionable.  For the $\Delta(1232)$, 
however, there is general agreement about the resonance 
width, mass, pole position, and residue~\cite{PDG}. Here, 
as the P$_{33}$ partial wave is elastic, a Breit-Wigner 
fit accurately describes the results of our more involved 
analysis. 

As mentioned, in fitting the electroproduction database, 
we fix the $Q^2$ = 0 point based on our fits to pion
photoproduction. The photoproduction multipoles can be
parametrized using the form
\begin{equation}
M = ({\rm Born} + \alpha_B )(1 + i T_{\pi N} )
  + \alpha_R T_{\pi N} + {\rm higher \; terms},
\end{equation}
\begin{figure}[htbp]
      \includegraphics[height=0.5\textwidth, angle=90]{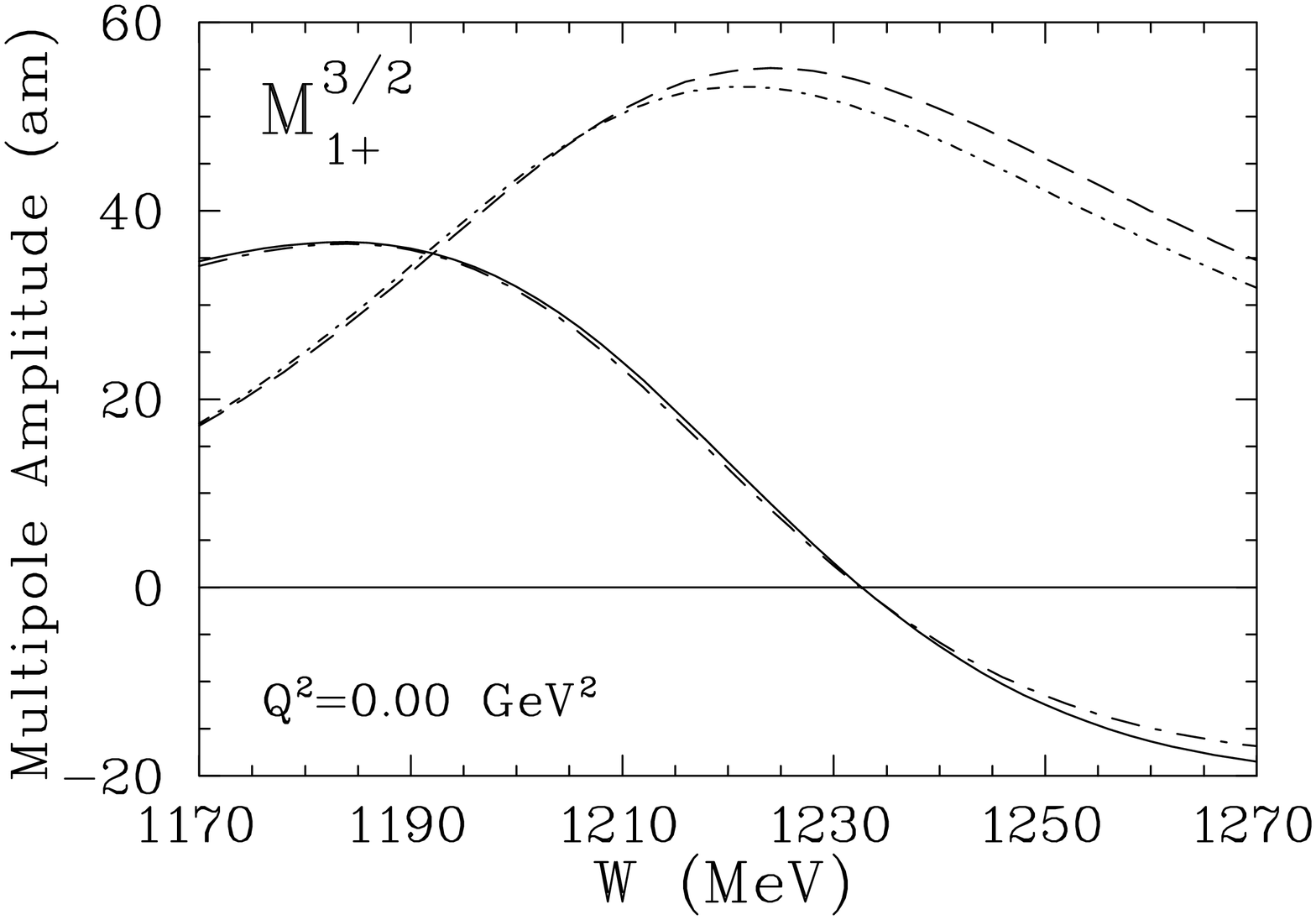}\hfill
      \includegraphics[height=0.5\textwidth, angle=90]{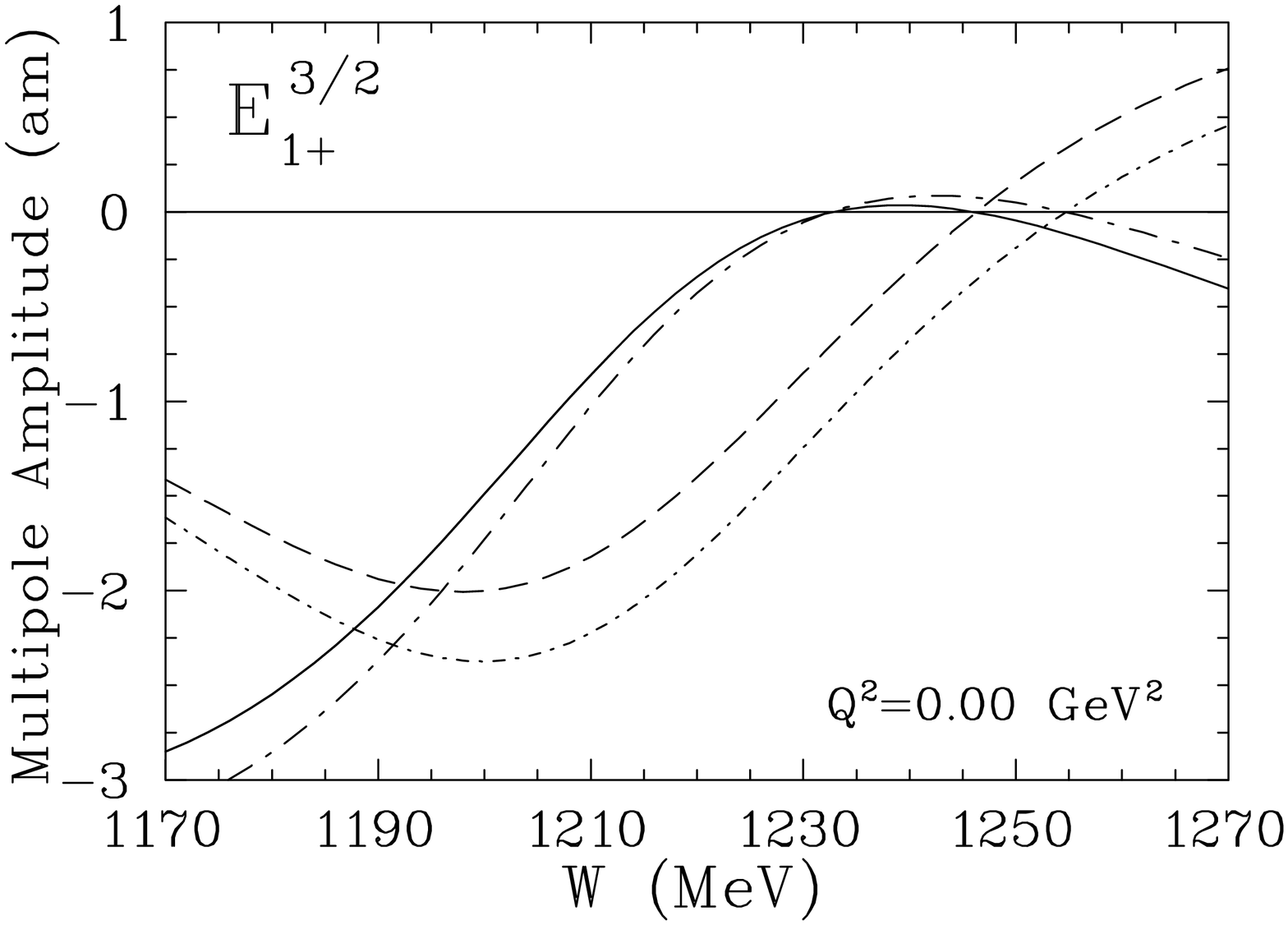}
      \caption{Partial-wave $P_{33}$ amplitudes around
      the $\Delta(1232)$ for $Q^2=0.00~GeV^2$. Magnetic
      ($M_{1+}^{3/2}$) and electric ($E_{1+}^{3/2}$)
      multipoles.  Solid (dashed) curves give the real
      (imaginary) parts of amplitudes corresponding to
      the pion phoproduction SP06
      solution~\protect\cite{gw_photo}.  The MAID05
      solution~\protect\cite{maid} is plotted with
      long dash-dotted (real part) and short
      dash-dotted (imaginary part) lines.}
      \label{fig:g5}
\end{figure}
containing the Born terms and phenomenological pieces 
($\alpha$) maintaining the correct threshold behavior and 
Watson's theorem below the two-pion production threshold. 
The $\pi N$ T-matrix ($T_{\pi N}$) connects each multipole 
to structure found in the elastic scattering analysis. In 
the $\Delta(1232)$ resonance region, the influence of 
channels beyond $\pi N$ is small for most partial waves 
and it is common to drop the $\alpha_B$ and higher order 
pieces. We retain the $\alpha_B$ term to account for
non-pole K-matrix contributions beyond the simple Born terms. 
\begin{figure}[htbp]
      \includegraphics[height=1.0\textwidth, angle=90]{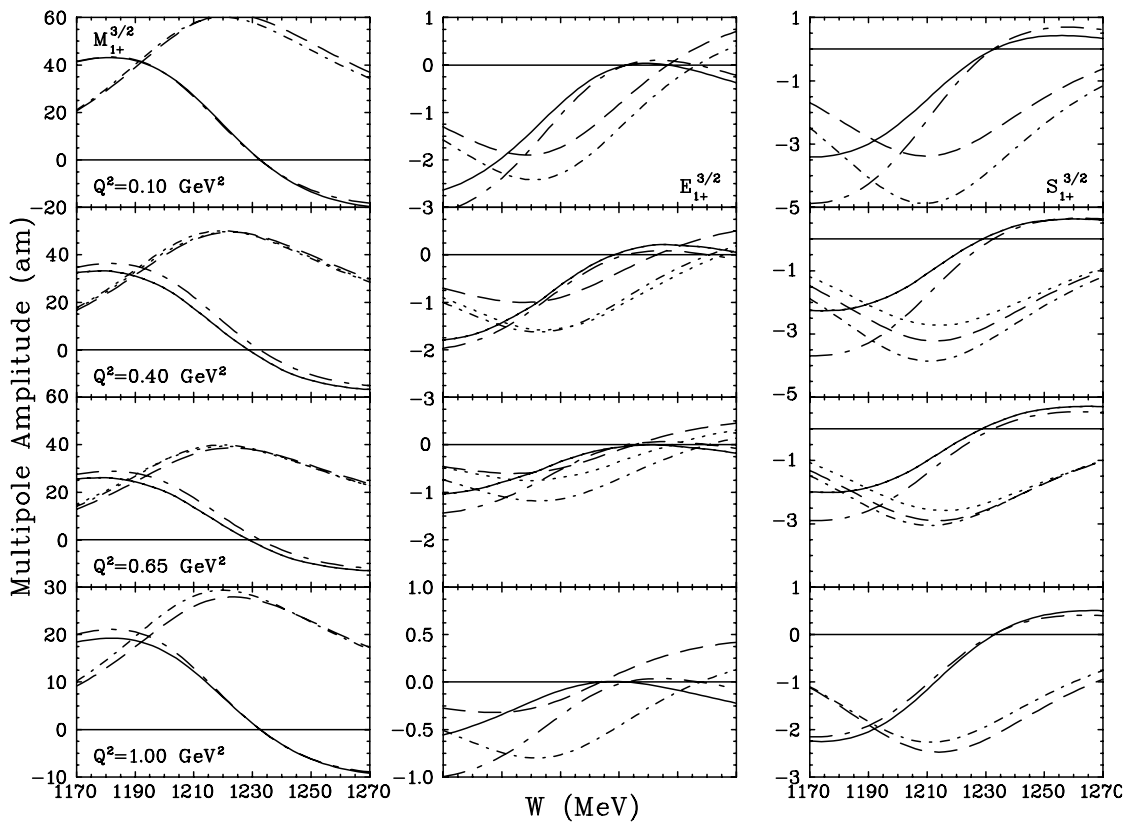}
      \caption{Magnetic ($M_{1+}^{3/2}$), electric
      ($E_{1+}^{3/2}$), and longitudinal
      ($S_{1+}^{3/2}$) $P_{33}$ multipoles for
      $Q^2<1.50~GeV^2$.  The JM05
      results~\protect\cite{CLAS} is plotted with
      dotted (real and imaginary parts) lines
      for $Q^2=0.40$ and $0.65~GeV^2$.  Notation
      as in Fig.~\protect\ref{fig:g5}.}
      \label{fig:g6}
\end{figure}
\begin{figure}[htbp]
      \includegraphics[height=1.0\textwidth, angle=90]{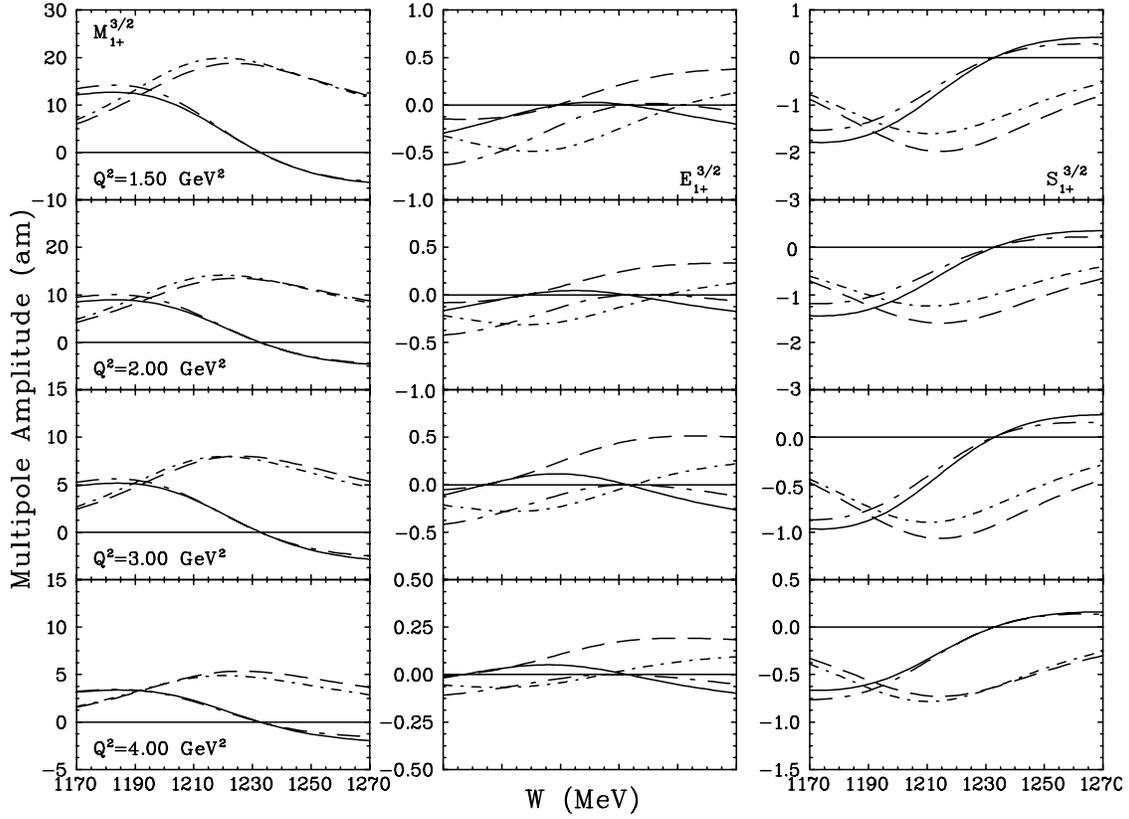}
      \caption{Magnetic, electric, and longitudinal
      $P_{33}$ multipoles for $Q^2>1.00~GeV^2$.
      Notation as in Fig.~\protect\ref{fig:g5}.} 
      \label{fig:g7}
\end{figure}

At non-zero $Q^2$, the Born terms have built-in $Q^2$ 
dependence $-$ other terms have been modified by a 
phenomenological factor
\begin{equation}
f(Q^2) = { k \over {k(Q^2=0)}} {1\over {1 + Q^2/0.7}} 
       e^{-\Lambda Q^2} 
       \left( 1 + Q^2 ( a + b \left[ {W \over W_R} - 
       1 \right] + c Q^2 ) \right)   
\end{equation}
where $k$ is the photon CM momentum, $\Lambda$ is a 
universal cutoff factor, and the constants $a$, $b$, and 
$c$ are searched for each multipole.  The $W$-dependent 
term is included to account for any residual energy 
dependence, is purely phenomenological, and was found to 
significantly improve fits. Note that this term is 
constructed to given zero contribution both at the $Q^2$=0 
(photoproduction) point and at the resonance position.
\begin{figure}[htbp]
      \includegraphics[height=0.5\textwidth, angle=90]{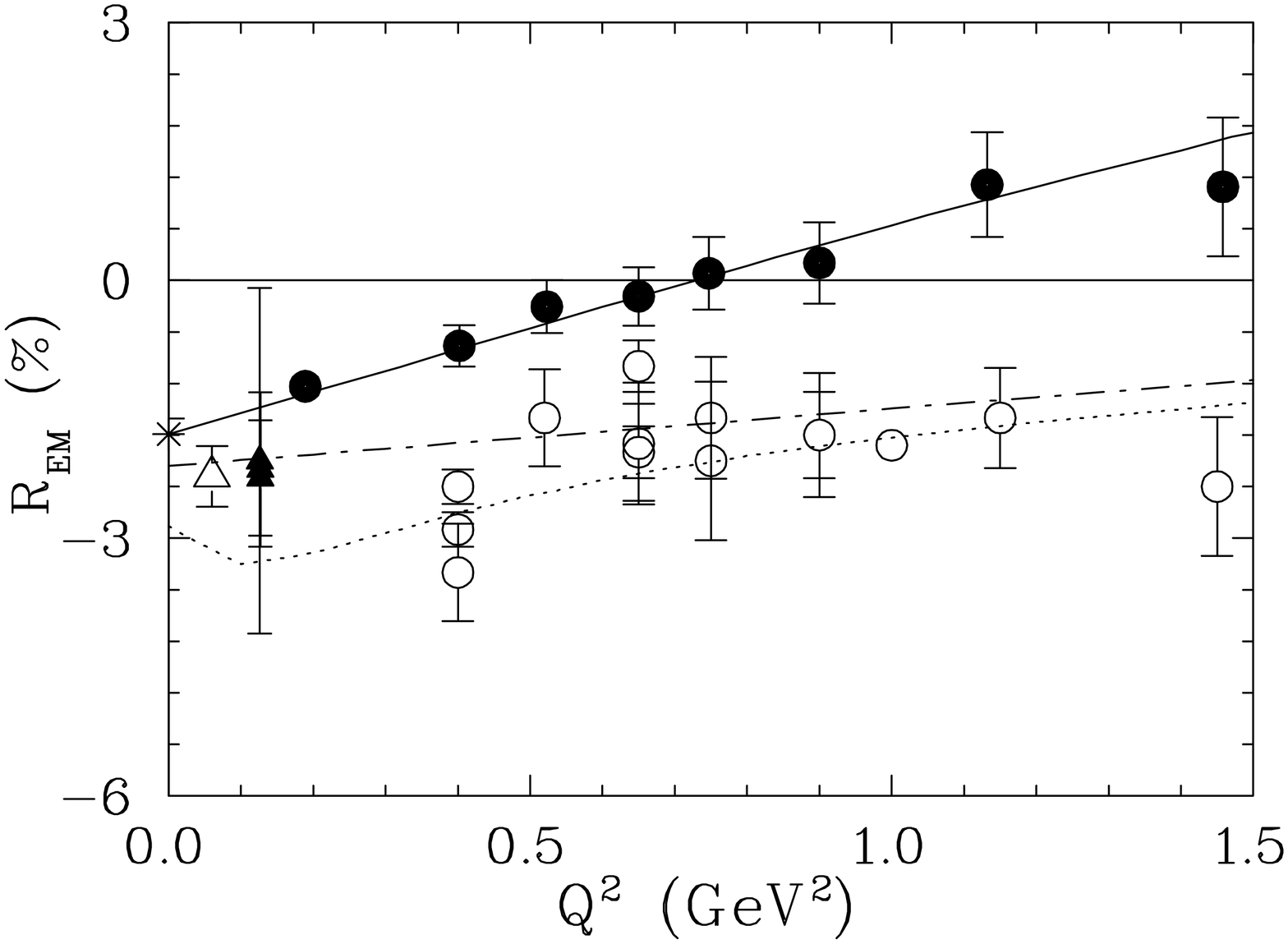}\hfill
      \includegraphics[height=0.5\textwidth, angle=90]{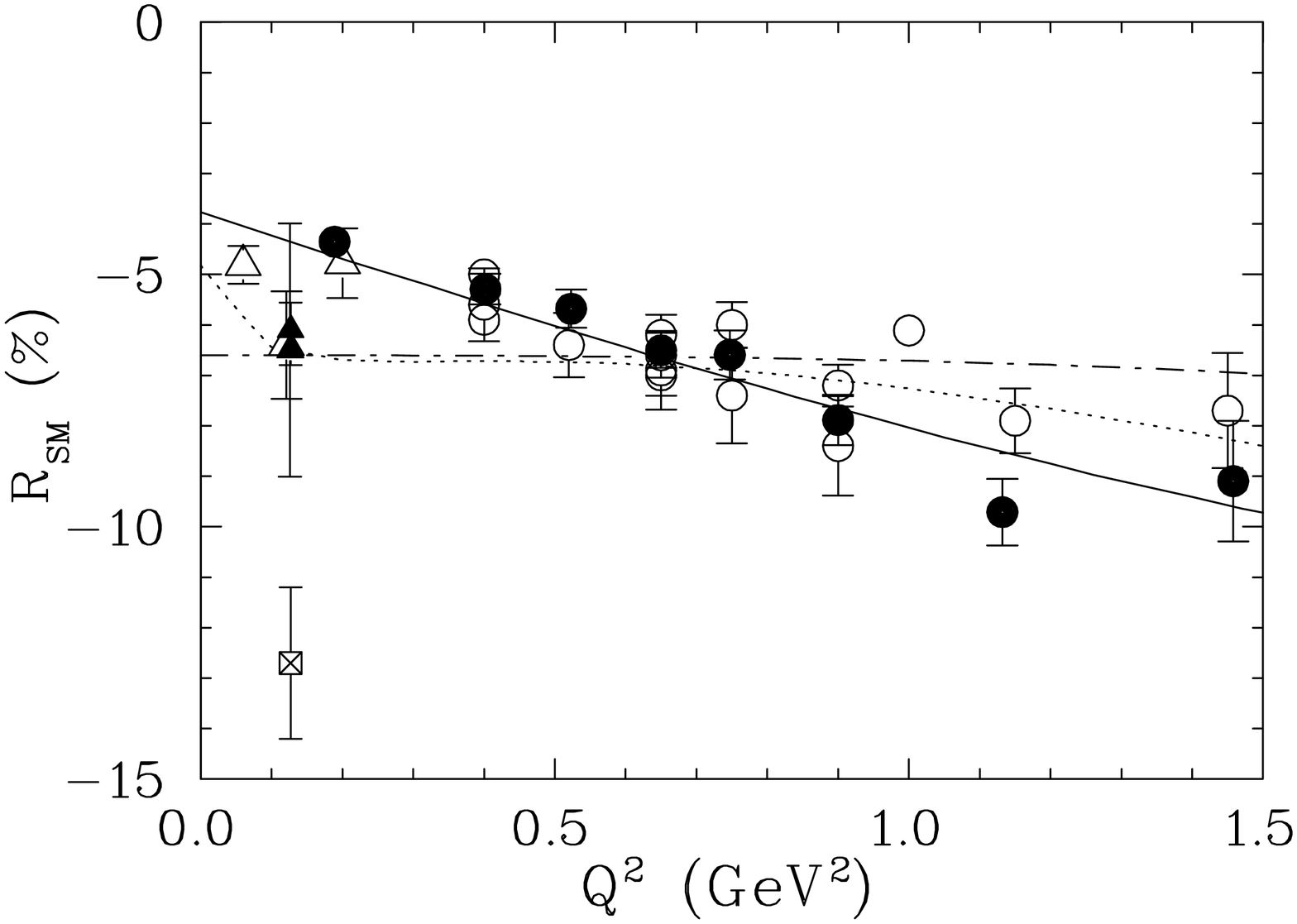}
      \caption{$R_{EM}$ and $R_{SM}$ ratios vs $Q^2$.
      Values were extracted from our fixed $Q^2$ analyses
      starting from the global fit (filled circles).
      Results from  JLab~\protect\cite{jlab} (open
      circles), MAMI-B~\protect\cite{mainz} (open
      triangles), MIT-Bates~\protect\cite{bates} (filled
      triangles), and ELSA~\protect\cite{elsa}
      (open square with cross) are given.  At $Q^2 = 0$,
      $R_{EM} = -1.79\pm 0.18~\%$ determined from pion
      photoproduction PWA~\protect\cite{gw_photo} is
      shown (star).  The solid curve gives our global
      (energy dependent) best-fit results.  The long
      dash-dotted (MAID05) and dotted (DMT) curves are
      from Refs.~\protect\cite{maid,DMT},
      respectively.} \label{fig:g8}
\end{figure}

In Fig.~\ref{fig:g5}, we compare the SAID and MAID results
for electric and magnetic multipoles connected to the 
$\Delta (1232)$ resonance. At the resonance position, the 
$R_{EM}$ ratio is essentially given by a ratio of the
imaginary parts of these multipoles. The large magnetic
multipole is not significantly different in these two
analyses (the agreement is even closer for the SAID and 
DMT multipoles~\cite{DMT}). Differences for the electric 
multipole are much larger. 

In Figs.~\ref{fig:g6} and~\ref{fig:g7}, we see the same 
trend continued for non-zero values of $Q^2$.  At the
resonance position, the magnetic ($M_{1+}^{3/2}$) 
multipoles remain fairly consistent; the $E_{1+}^{3/2}$ 
multipoles differ significantly. 

One point of continuing interest has been the $Q^2$
variation of the ratio $R_{EM}$ - a quantity which should 
\begin{figure}
      \includegraphics[height=0.5\textwidth, angle=90]{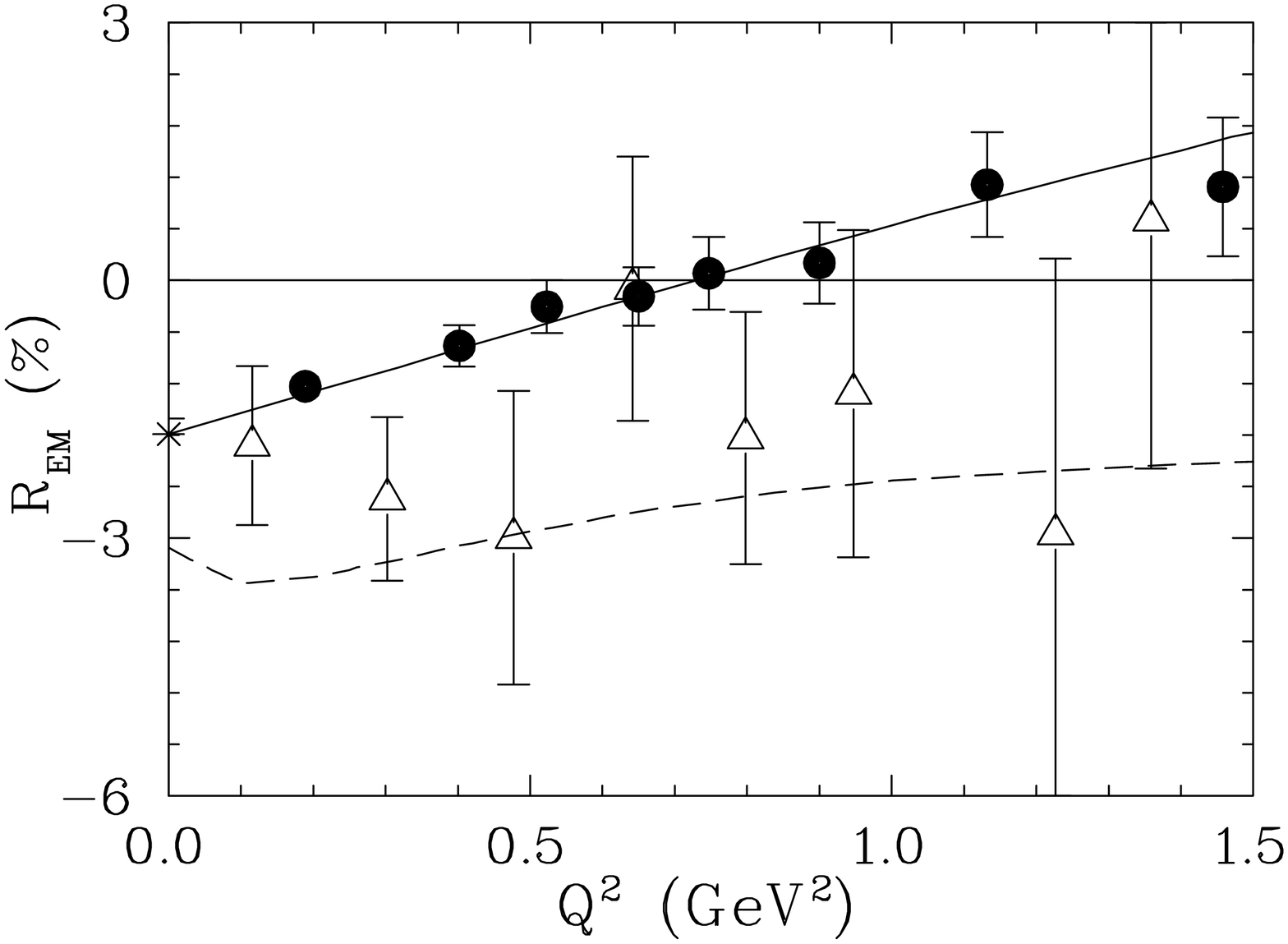}\hfill
      \includegraphics[height=0.5\textwidth, angle=90]{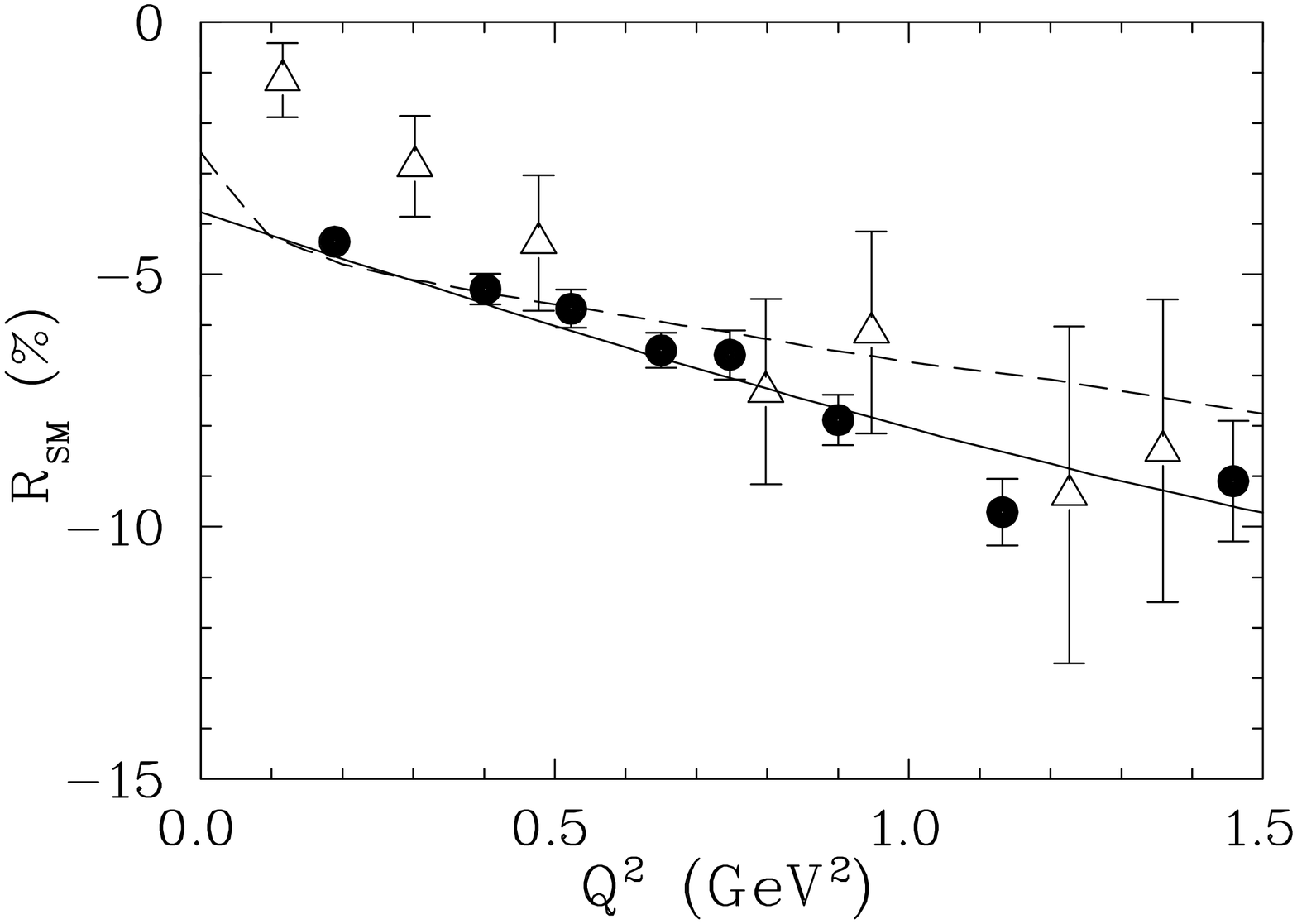}
      \caption{$R_{EM}$ and $R_{SM}$ ratios vs $Q^2$.
      Recent lattice~\protect\cite{lattice} and
      previous S-L~\protect\cite{S-L} calculations
      shown as open triangles and long dashed line,
      respectively.} \label{fig:g9}
\end{figure}
tend to unity for suffiently high $Q^2$.  In the SAID fits, 
a cross-over from negative to positive values of $R_{EM}$ 
has been found, though the exact cross-over point has 
tended to shift as our database has expanded.  The most 
recent results are shown in Fig.~\ref{fig:g8} compared to 
a number of models and single-$Q^2$ fits to data.  Here, 
we also plot the ratio $R_{SM}$, for which there is at 
least qualitative agreement up to about 4~GeV$^2$. (The 
incorporation of high-$Q^2$ data is not complete, so there 
could yet be changes.)  While our present curve for $R_{EM}$ 
appears at odds with other determinations, it is not ruled 
out by recent lattice data~\cite{lattice}, displayed in 
Fig.~\ref{fig:g9}.

We plan to continue these fits, incorporating all available
electroproduction data, and modifying our fitting procedure 
as necessary.  Useful comparisons will require those involved 
in this effort to make available all amplitudes obtained in 
any new determination of $R_{EM}$ and $R_{SM}$. 


\begin{theacknowledgments}
This work was supported in part by the U.~S.~Department of
Energy Grant DE--FG02--99ER41110 and by funding from 
Jefferson Laboratory.
\end{theacknowledgments}

\end{document}